\documentclass[twocolumn,twoside,slac_two]{revtex4}
\usepackage{graphicx}
\usepackage{fancyhdr}
\begin{document}
\title{The origin of the prompt GRB spectrum}

\author{F. Daigne\footnote{Institut Universitaire de France}}
\author{Z. Bosnjak\footnote{Present: CEA Saclay, DSM/IRFU/Service d'Astrophysique -- 91191 Gif-sur-Yvette, France}}
\affiliation{Institut d'Astrophysique de Paris -- UMR 7095 Universit\'{e} Pierre et Marie Curie-Paris 06; CNRS \\ 98 bis bd Arago, 75014 Paris, France}
\author{G. Dubus}
\affiliation{Laboratoire d'Astrophysique de Grenoble -- UMR 5571 Universit\'{e} Joseph Fourier; CNRS\\ BP 53, 38041 Grenoble, France}

\begin{abstract}
Using a detailed model of the internal shock phase, we discuss the origin of the prompt emission in gamma-ray bursts. We focus on the identification of the dominant radiative process (Fermi-GBM range) and propose an explanation for some features observed by Fermi-LAT at high energy in some GRB lightcurves.
\end{abstract}
\maketitle

\thispagestyle{fancy}

\section{Introduction}
The physical origin of the prompt emission from Gamma-ray bursts (hereafter GRBs) is still unclear.  Due to the high variability
observed in the lightcurves, it is usually believed that it is produced from within the relativistic outflow (internal origin) and that 
the variability is due to the activity of the central engine \citep{sari:97}. Then the radiated energy can in principle be extracted from
three potential reservoirs : thermal, kinetic or magnetic energy. In the first case, radiation will occur at the photosphere. In the two other cases,
an extraction mechanism is needed. The extraction of the kinetic energy of the outflow can be obtained via shock waves propagating
within the outflow (internal shocks, \citet{rees:94}) that accelerate particles, which then radiate. The extraction of magnetic energy can be
achieved via magnetic dissipation due to the reconnection of the field lines \citep{lyutikov:03}. This leads again to particle acceleration.\\

The first fundamental
question is to identify which of these mechanisms is (are ?) at work in GRBs. An additional question is to understand which radiative processes are
responsible for the emission and the shape of the spectrum. 
In order to model the phenomenology of GRBs (lightcurves, spectrum, spectral evolution) both questions (energy extraction mechanism and
dominant radiative process) cannot be considered independently. Indeed the radiated spectrum cannot be computed without estimating the
physical conditions in the emitting region.  \\

In this contribution we focus on the identification of the dominant radiative process in the prompt emission from GRBs assuming that
it is produced by shock-accelerated electrons in internal shocks occuring above the photospheric radius of the outflow. It assumes that
the magnetization of the outflow at large distance of the central source is weak ($\sigma < 1$), otherwise shock waves would not propagate within the outflow.
It also assumes that the photospheric emission is weak and hidden by the non-thermal radiation from internal shocks. Due to the low efficiency
of the latter, it implies either that the outflow is ejected from the innermost region of the central engine ($r_0 \le 10^{6}\ \mathrm{cm}$) or that 
the outflow is initially highly magnetized ($\sigma > 10$) and that 
the acceleration is due to an efficient conversion of the magnetic energy into kinetic energy below the photosphere \citep{daigne:02}. \\

In such a situation, the two main radiative processes that may be responsible to build the optically thin non-thermal spectrum are synchrotron radiation or inverse Compton scatterings. This is discussed in Sect.~\ref{sec:rad}. Then we present in Sect.~\ref{sec:is} some results obtained from a model of the internal shock phase that
includes a detailed treatment of the dynamics and the radiation. Sect.\ref{sec:discuss} is the conclusion.

\begin{figure}[t]
\centering
\includegraphics[width=0.45\textwidth]{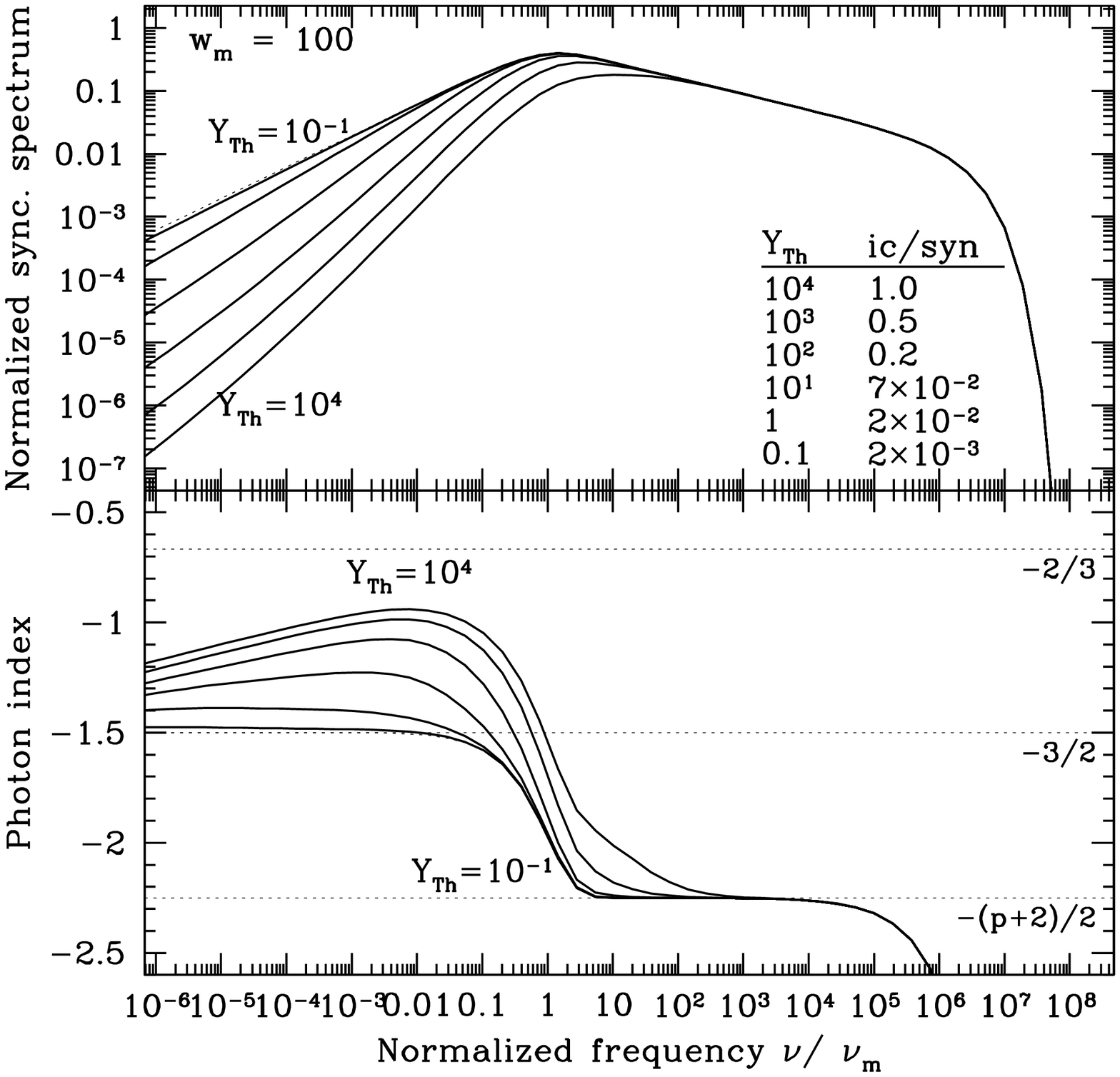}
\caption{\textbf{Steepening of the synchrotron spectrum due to inverse Compton scatterings in Klein-Nishina regime} (from \citet{daigne:10}). Using a detailed radiative code, we compute the synchrotron spectrum for a fixed value $w_\mathrm{m}=100$ of the parameter measuring the importance of Klein-Nishina corrections and for a an increasing value of the parameter $Y_\mathrm{Th}$ measuring the importance of inverse Compton scatterings (see text). The low-energy photon index is steepening from $-3/2$ for low $Y_\mathrm{Th}$ to $\sim -1$ for high $Y_\mathrm{Th}$ (bottom panel). The inverse Compton process remains energetically sub-dominant as indicated by the inserted table in the top panel.} \label{fig:slope1}
\end{figure}

\section{\label{sec:rad}What is the dominant radiative process during the prompt GRB emission ?}
There are mainly two possibilities for the origin of the soft $\gamma$-rays observed in GRBs : either they are directly emitted
by synchrotron radiation from highly relativistic electrons or they are produced by inverse Compton
scatterings of soft synchrotron photons by relativistic electrons (SSC). Observations seem now to favor the synchrotron scenario.
Indeed, in the SSC case, the final shape of the spectrum is mainly determined by the value of the parameter $Y_\mathrm{Th}\sim \epsilon_\mathrm{e}/\epsilon_\mathrm{B}$, where $\epsilon_\mathrm{B}$ (resp. $\epsilon_\mathrm{e}$) is the fraction of the internal energy available in the shocked region which 
is transfered in the magnetic field (resp. the accelerated electrons). Depending on the value of $Y_\mathrm{Th}\sim \epsilon_\mathrm{e}/\epsilon_\mathrm{B}$, it is expected that either the synchrotron component at low energy is dominant ($Y_\mathrm{Th}<1$), leading to a bright prompt optical
emission which is not observed, or that the dominant component is the second inverse Compton peak ($Y_\mathrm{Th}>1$), due to second scatterings
(third and other scatterings are suppressed by Klein-Nishina effects). This predicts a strong component that does not seem to be detected
by Fermi-LAT \citep{piran:09,bosnjak:09}. In addition, the peak energy of the soft $\gamma$-ray component is highly sensitive to the physical
conditions in the emitting region in the SSC scenario, as the typical energy of inverse Compton photons in Thomson regime is proportional
to $B \Gamma_\mathrm{m}^4$, where $B$ is the magnetic field and $\Gamma_\mathrm{m}$ the typical Lorentz factor of shock-accelerated
electrons. This leads to a spectral evolution in pulses in GRB lightcurves which is too rapid compared to observations \citep{daigne:98}. In the synchrotron
scenario, these two problems do not appear : the spectral evolution is slower and there is no bright component expected in the Fermi-LAT
range as even the first inverse Compton scatterings occur in Klein-Nishina regime. Indeed high electron Lorentz factors are needed to produce
synchrotron photons in the soft $\gamma$-ray range and therefore, the parameter $w_\mathrm{m}=\Gamma_\mathrm{m} h \nu'_\mathrm{m}/ m_\mathrm{e} c^2$, where $\nu'_\mathrm{m}$ is the synchrotron frequency of electrons at $\Gamma_\mathrm{m}$, is large (Klein-Nishina corrections become important for $w_\mathrm{m}>1$).\\

There is however a well known problem in the synchrotron scenario, related to the low-energy photon index measured in GRB spectra \citep{ghisellini:00}. To reproduce both the high variability and the huge luminosity in GRB lightcurves, it is necessary that the relativistic electrons are radiatively efficient, i.e. that their radiative timescale is shorter than the dynamical
timescale. When only synchrotron radiation is considered, this condition is equivalent to $\Gamma_\mathrm{c}<\Gamma_\mathrm{m}$, where $\Gamma_\mathrm{c}$ is defined as the Lorentz factor of electrons having a synchrotron time scale equal to the dynamical timescale. In this fast-cooling regime, the low-energy photon index below the peak of $\nu F_\nu$ is $\alpha=-3/2$ \citep[see e.g.][]{sari:98}, in contradiction with measured values which are centered around $-1$ \citep{preece:00}. \\
\begin{figure}[t]
\centering
\includegraphics[width=0.45\textwidth]{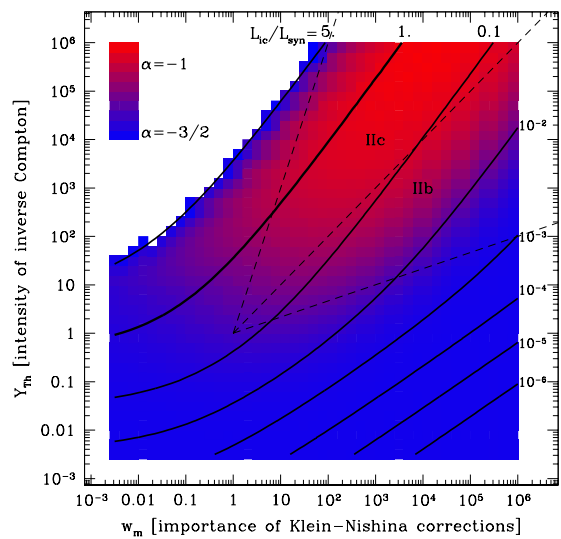}
\caption{\textbf{The low-energy index $\alpha$ of the fast-cooling synchrotron spectrum} (from \citet{daigne:10}). Using radiative calculations including only synchrotron radiation and inverse Compton scatterings (assuming $\Gamma_\mathrm{c}\ll \Gamma_\mathrm{m}$), the low-energy photon index $\alpha$ is plotted as a function of $w_\mathrm{m}$ and $Y_\mathrm{Th}$. For high values of these two parameters it becomes as steep as $-1$. Lines of constant ratio of the luminosity radiated in the inverse Compton component over the luminosity radiated in the synchrotron component are also plotted and show that inverse Compton scatterings are usually energetically sub-dominant. Note that this ratio should be even lower when including $\gamma\gamma$ annihilation. } \label{fig:slope2}
\end{figure}
\begin{figure*}[t]
\centering
\includegraphics[width=0.45\textwidth]{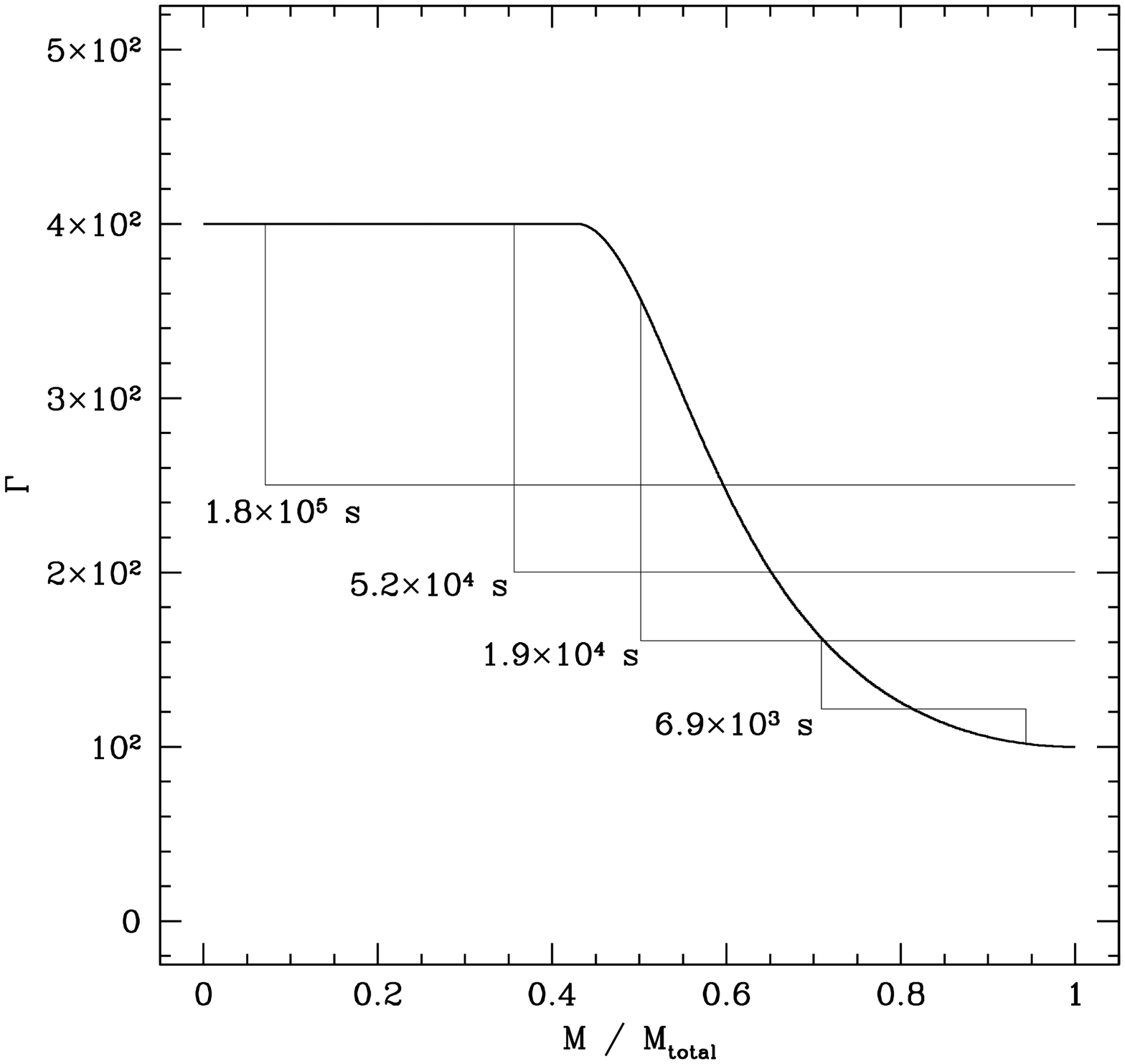}
\includegraphics[width=0.45\textwidth]{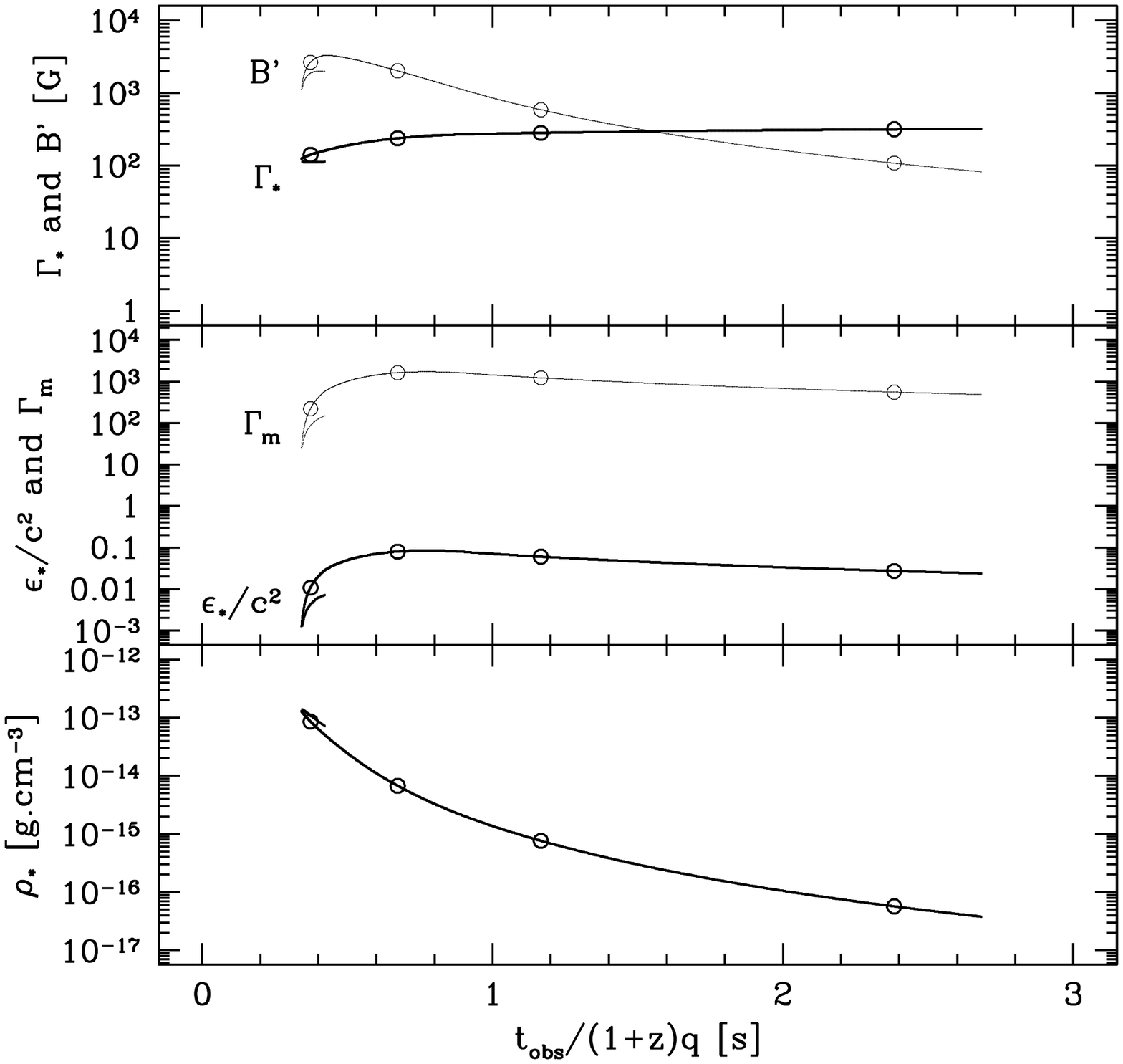}
\caption{\textbf{An example of internal shock dynamics} (from \citet{daigne:98}). Left: evolution of the Lorentz factor distribution within the outflow in Lagrangian coordinates. The initial conditions (relativistic ejection by the central engine) are described in the text. The formation of two shocks appears clearly. The first ones disappears rapidly as it reaches the front edge and the second shock crosses the whole ejecta. Right: the physical conditions in the shocked region are plotted as a function of the observer time $t_\mathrm{obs}/(1+z)=t-R/c$ : Lorentz factor $\Gamma_*$, comoving density $\rho_*$ and specific internal energy density $\epsilon_*$. In addition, the evolution of the magnetic field $B'$ and the electron Lorentz factor $\Gamma_\mathrm{m}$ is also indicated for a given set of microphysics parameters.} \label{fig:dyn}
\end{figure*}

A typical GRB spectrum is well fitted by a phenomenological function introduced by \citet{band:93}. Assuming a high-energy photon index $\beta=-2.3$, it is easy to calculate
that it is necessary to remove only $\sim 20\%$ of the energy radiated in the synchrotron component to move $\alpha$ from $-3/2$ to $-1$. This means that in principle,
a sub-dominant process should be enough to solve the problem. It is tempting to suggest that inverse Compton scatterings in Klein-Nishina regime play this role \citep{derishev:01}. Indeed, Klein-Nishina limitations are more important for the scatterings of high-energy photons and therefore the probability to be scattered increases when the photon energy decreases, leading to a steepening of the synchrotron spectrum. In Thomson regime ($w_\mathrm{m}<1$), all photons have equal probability to be scattered and the synchrotron shape is unaffected. This effect is illustrated in Fig.~\ref{fig:slope1}. The spectra shown in this figure have been computed using a radiative code
developed for the purpose of GRB studies \citep{bosnjak:09},
that solves simultaneously the time-evolution of electrons and photons. For clarity, the calculations done in Fig.~\ref{fig:slope1} include only synchrotron radiation
and inverse Compton scatterings. Figure~\ref{fig:slope2} illustrates the evolution of the low-energy photon index $\alpha$ in the parameter space $Y_\mathrm{Th}$ (importance of inverse Compton scatterings) versus $w_\mathrm{m}$ (importance of Klein-Nishina corrections). These numerical results agree well with the simplified semi-analytical model developped by \citet{nakar:09}. It appears then that there is a large region of the parameter space that allows steep slopes ranging from $-3/2$ to $-1$. Lines of constant ratio $L_\mathrm{ic}/L_\mathrm{syn}$ are also plotted in Fig.~\ref{fig:slope2} and show that inverse Compton scatterings remain sub-dominant in most situations.\\

These results, which will be discussed in more details in a paper to be submitted \citep{daigne:10}, offer a nice possibility to reconcile synchrotron radiation with GRB observed spectra. However, to go further and check if the conditions necessary for steep slopes can indeed be reached in GRBs, one needs to make some assumptions regarding the physical mechanism responsible for the extraction of the energy in the prompt GRB phase. This is done in the next section in the framework of the internal shock model.

\begin{figure*}[t]
\centering
\includegraphics[width=0.45\textwidth]{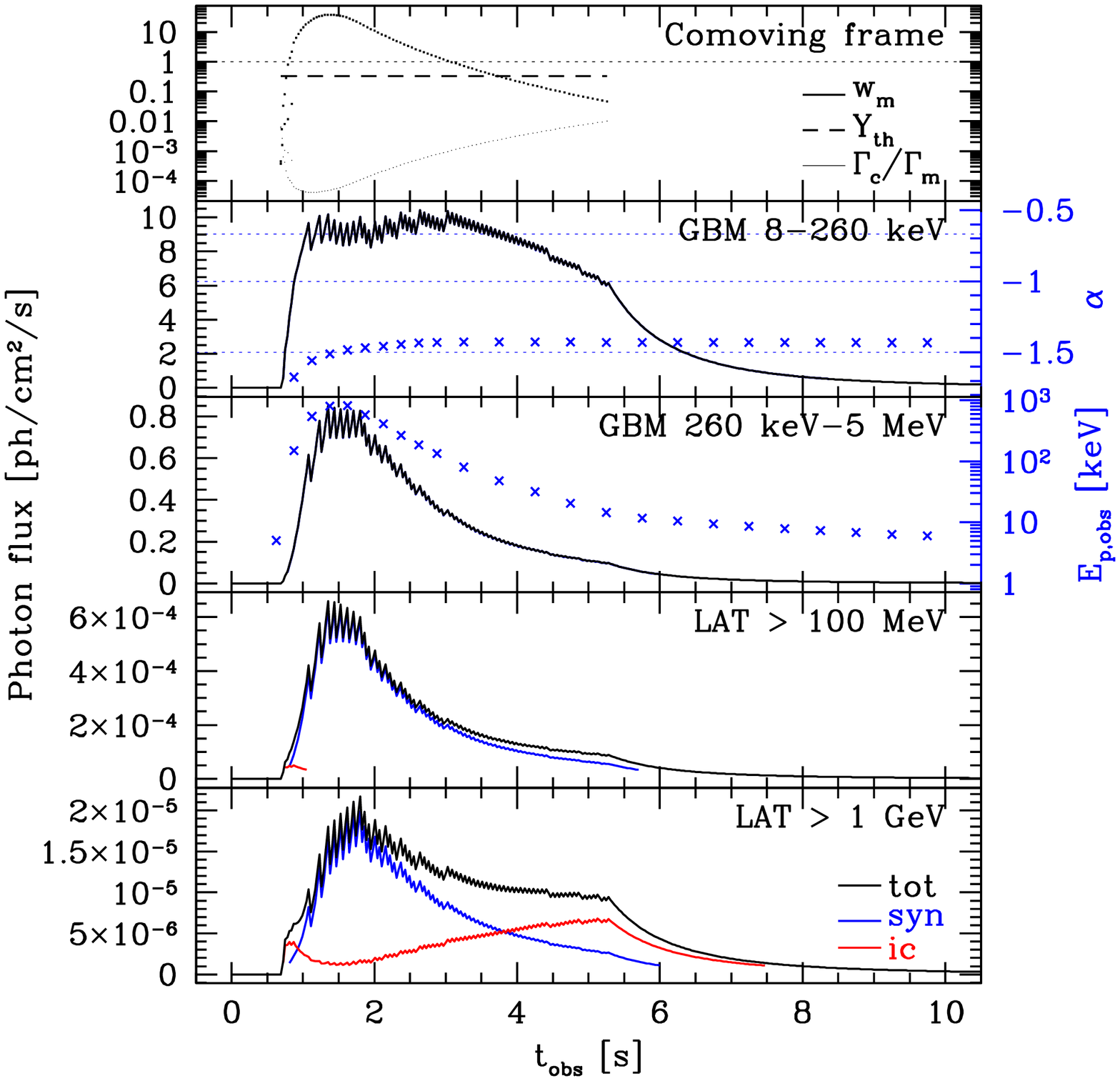}
\includegraphics[width=0.45\textwidth]{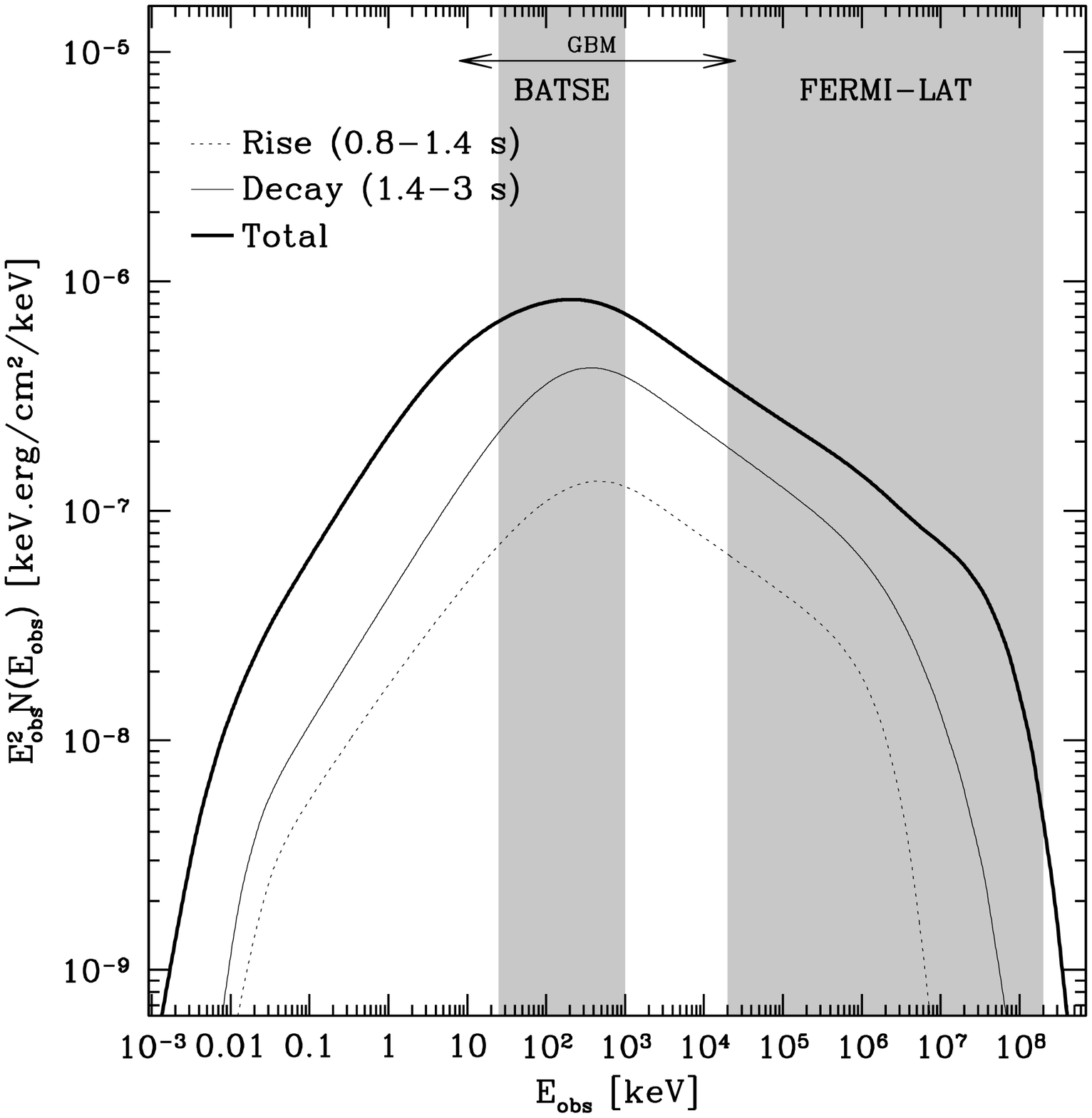}\\
\includegraphics[width=0.45\textwidth]{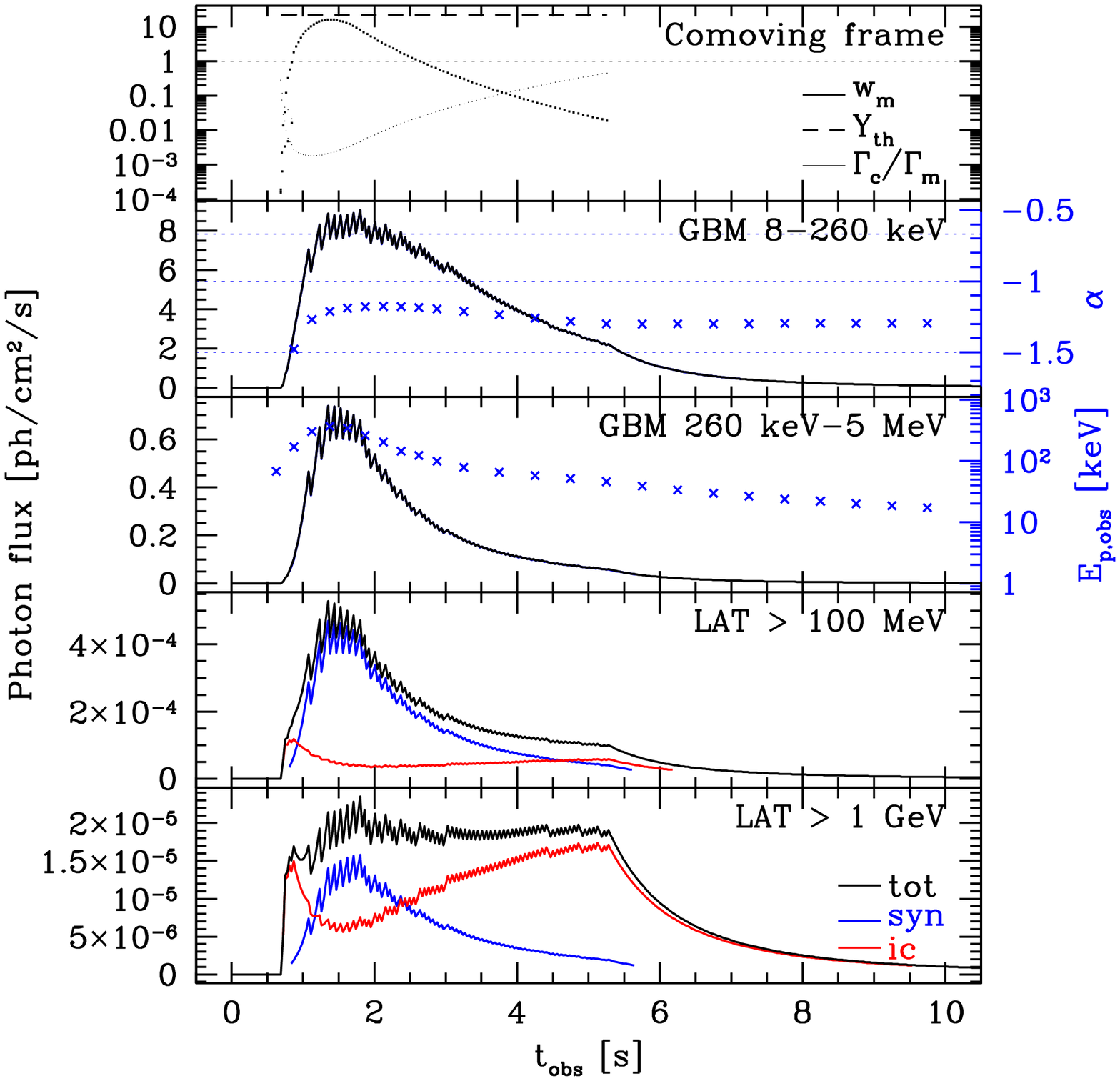}
\includegraphics[width=0.45\textwidth]{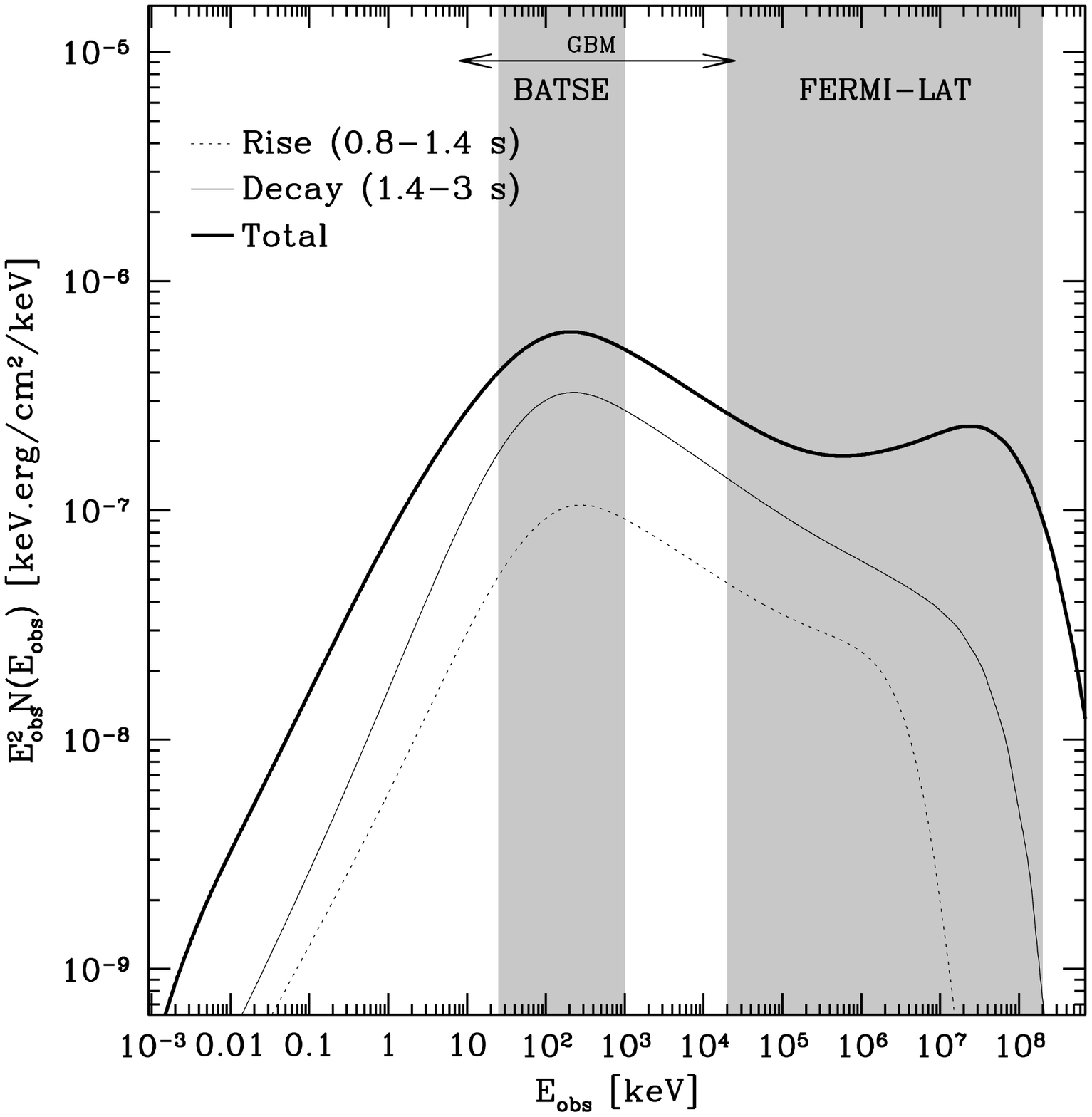}\\
\caption{\textbf{Two examples of a synthetic single pulse burst} (from \citet{bosnjak:09,daigne:10}). Using the simulation presented in Fig~\ref{fig:dyn}, lightcurves (left) and spectra (right) are computed using two different sets of microphysics parameters (see text) : $\epsilon_\mathrm{e}=\epsilon_\mathrm{B}=1/3$ and $\zeta=3\times 10^{-3}$ (top) and $\epsilon_\mathrm{e}=1/3$, $\epsilon_\mathrm{B}=5\times 10^{-3}$ and $\zeta=2\times 10^{-3}$. In addition, the top panel above the lightcurves shows the evolution of $w_\mathrm{m}$, $\Gamma_\mathrm{c}/\Gamma_\mathrm{m}$ and $Y_\mathrm{Th}$ in the shocked region.} \label{fig:examples}
\end{figure*}

\section{\label{sec:is}The expected emission from internal shocks}
We have developped a detailed model of the emission from internal shocks. The dynamics of the shock waves is followed using the model by \citet{daigne:98}. It allows to compute the evolution of the physical conditions in shocked regions within the outflow. We use then a standard parametrization to evaluate the magnetic field and the distribution of relativistic electrons in these regions : $\epsilon_\mathrm{e}$ and $\epsilon_\mathrm{B}$ as defined earlier. In addition we assume that only a fraction $\zeta$ of the electrons are accelerated. The radiation produced at a given instant during the propagation of a shock wave is computed in the comoving frame using our radiative code, including the most relevant processes : synchrotron radiation and self-absorption, inverse Compton scatterings, $\gamma\gamma$ annihilation and adiabatic cooling. Finally we compute the observed flux by integrating these elementary contributions over equal-arrival time surfaces. This allows to produce synthetic GRBs with lightcurves in different channels and spectra in different time intervals, and to compare their properties with observations.\\

Figure~\ref{fig:dyn} shows the dynamics of internal shocks in a simple example that leads to a single pulse burst. It has to be considered as a building block for more complex lightcurves. The central engine is ejecting relativistic matter during 2 s, with a constant kinetic energy flux $\dot{E}=5\times 10^{53}\ \mathrm{erg/s}$ and a Lorentz factor which increases from 100 at the beginning of the ejection to 400 at the end. This ejection leads to the propagation of internal shocks : the properties in the shocked medium are also plotted in the same figure. Note that some of these properties evolve strongly (see for instance the magnetic field). This leads to some spectral evolution in the radiated pulse, as shown in Fig~\ref{fig:examples} where the lightcurve and the spectrum of the emitted GRB are plotted for two different sets of microphysics parameters (high or low $\epsilon_\mathrm{B}$).\\

 These two examples show several interesting features. The dominant component in the spectrum is the synchrotron component that peaks in the soft $\gamma$-ray range (Fermi-GBM). The evolution of the physical conditions in the shocked region leads to a decrease of the peak energy of this component with time. The expected spectral evolution is therefore reproduced, at least qualitatively : hard-to-soft evolution in the pulse, the pulse peaks earlier at higher energy, the width of the pulse increases at low energy, etc. In addition to the peak energy, the low-energy index is also evolving (due to the fact that $w_\mathrm{m}$ decreases with time). For the low $\epsilon_\mathrm{B}$ case (high $Y_\mathrm{Th}$), as expected from the previous section, it is steeper than $-3/2$  for most of the pulse ($\sim -1.2\to-1.1$ for this example). Due to the high values of $w_\mathrm{m}$, the additional inverse Compton component is initially limited by Klein-Nishina effects and is very weak. However, this limitation is reduced at later times and in the case with a low  $\epsilon_\mathrm{B}$ (high $Y_\mathrm{Th}$), this leads to the delayed emergence of a well identified additional component in the spectrum at high-energy (Fermi-LAT range).   Such a variable additional component at high-energy, delayed compared to the GBM lightcurves, is observed in a good fraction of LAT bursts (see e.g. \citet{abdo:09}). Note that there is also a weaker precursor in the GeV range in the second example. This is due to the assumed shape of the initial distribution of the Lorentz factor in the outflow (see Fig.~\ref{fig:dyn}) that leads to an internal shock which is only mild initially and become more violent later. This precursor disappears in simulations done assuming a steeper variation of the Lorentz factor in the initial ejecta. On the other hand, the delayed GeV emission at the end of the pulse is a generic feature as it is directly related to a propagation effect (especially the decrease of the magnetic field with distance). All these calculations are done assuming constant microphysics parameters during the shock propagation, which is not necessarily the case. Then depending on the evolution of these parameters it could either amplify or reduce the behaviour presented in these two examples.

\section{\label{sec:discuss}Discussion and conclusion}
We have presented a detailed model of the emission from internal shocks propagating within a relativistic outflow and applied it to the prompt emission from GRBs. In this framework, recent observations, including the detection of a few GRBs by Fermi-LAT, strongly favor the situation where the dominant radiative process is synchrotron radiation. We show that inverse Compton scatterings in Klein Nishina regime should occur and can steepen the low-energy photon index of the synchrotron component from $-3/2$ to $-1$, which is in better agreement with observations. In addition, we show that the model has the capacity to reproduce the spectral evolution observed in pulses in the soft $\gamma$-ray range (GBM) and predict, for low values of $\epsilon_\mathrm{B}$ the presence of a variable additional component in the GeV range which emerges with a delay due to an initial limitation of inverse Compton scatterings by the Klein-Nishina effect. This scenario offers a possible explanation to several features observed in the lightcurve of the bursts detected by Fermi. It also puts interesting constraints on the microphysics parameters and therefore on the physics of shock acceleration. 

\bigskip 
\bibliography{daignef}
\end{document}